\documentstyle[emulateapj,epsfig]{article}

\begin{document}
\submitted{ApJ in press}

\title{The Expected Redshift Distribution of Gamma-Ray Bursts}

\author{Volker Bromm and Abraham Loeb}

\affil{Astronomy Department, Harvard University, 60 Garden Street,
  Cambridge, MA 02138; vbromm@cfa.harvard.edu, aloeb@cfa.harvard.edu}

\begin{abstract} 
We predict the redshift distribution of Gamma-Ray Bursts (GRBs) assuming
that they trace the cosmic star formation history. We find that a fraction
$\ga 50\%$ of all GRBs on the sky originate at a redshift $z\ga 5$, even
though the fraction of the total stellar mass formed by $z\sim 5$ is only
$\sim 15$\%.  These two fractions are significantly different because they
involve different cosmological factors when integrating the star formation
rate over redshift.  Hence, deep observations of transient events, such as
GRB afterglows or supernovae, provide an ideal strategy for probing the
high-redshift universe.  We caution, however, that existing or planned
flux-limited instruments are likely to detect somewhat smaller fractions of
high redshift bursts.  
For example, we estimate that the fraction of all 
bursts with redshifts $z\ga 5$ is $\sim$10\% in the case of the BATSE
instrument, and $\sim$25\% in the case of {\it Swift}.
We also show that the intrinsic distribution of GRB
durations is bimodal but significantly narrower and shifted towards shorter
durations than the observed distribution.
\end{abstract}


\section{Introduction}

Gamma-Ray bursts (GRBs) are the brightest electromagnetic explosions in the
universe (for a recent review, see Piran 2000).  Popular models for their
central engine divide into two main classes: (i) the collapse of a massive
star to a black hole (BH) (MacFadyen, Woosley, \& Heger 2001, and
references therein); (ii) the coalescence of a binary system involving a
neutron star (NS) and a BH or a NS as a companion (e.g.  Eichler et
al. 1989; Janka et al. 1999).  The observed association of long-duration
GRBs with star forming regions (Djorgovski et al. 2001c, and references
therein), and the possible supernova signatures in rapidly-decaying
afterglows (Bloom et al. 1999; Kulkarni et al. 2000; Reichart 2001) favors
the first class.  Both classes of models associate GRB progenitors with
compact objects (BH or NS) that are the end products in the evolution of
massive stars. Hence, the GRB formation history is expected to follow the
cosmic star formation history (Totani 1997, 1999; Wijers et al. 1998; Blain
\& Natarajan 2000) up to the highest redshifts ($z\sim 20$) at which the
first generation of stars may have formed (Barkana \& Loeb 2001). GRBs
might therefore provide an ideal probe of cosmic star formation at all
redshifts that in particular is unaffected by dust obscuration (e.g., Blain
\& Natarajan 2000; Porciani \& Madau 2001).  In fact, the top-heavy initial
mass function (IMF) predicted for the first stars (Bromm, Coppi, \& Larson
1999, 2002; Abel, Bryan, \& Norman 2000, 2002, Nakamura \& Umemura 2001)
favors massive stars which are the likely source of GRB progenitors.

GRB afterglows provide a unique probe of the high redshift universe (Lamb
\& Reichart 2000; Ciardi \& Loeb 2000). The bright, early optical-UV
luminosity of a GRB afterglow is expected to outshine its host galaxy, even
more so at high redshifts when the typical galaxies are less massive than
their present-day counterparts (Barkana \& Loeb 2001).  The broad-band
afterglow spectrum extends into the far UV and so the absorption features
imprinted on it by the intervening intergalactic medium (IGM) can be used
to infer the evolution of the neutral hydrogen fraction and the metal
abundance of the IGM during the epoch of reionization. In difference from
galaxies and quasars, which fade rapidly with increasing redshift due to
the increase in their luminosity distance, GRB afterglows maintain an
almost constant infrared flux with increasing redshift at a fixed time lag
after the GRB trigger in the observer frame (Ciardi \& Loeb 2000). This
follows from the cosmological time--stretching of the afterglow transient
(which is intrinsically brighter at earlier times) and from a favorable
$K$-correction in the afterglow spectrum.

The {\it Swift} satellite\footnote{See http://swift.gsfc.nasa.gov/},
planned for launch in 2003, is expected to localize roughly one GRB per
day. Sorting out the subset of all GRBs which originate at high redshifts
($z\ga 5$) would be of particular interest.  Observers may employ a simple
strategy for this purpose.  Photometric data from a small telescope should
be used at first to identify those GRBs which possess a Ly$\alpha$ trough
at a wavelength of $0.73\mu{\rm m}(1+z)/6$ due to absorption by the
IGM. Follow-up spectroscopy of those GRBs could then be done on a 10-m
class telescope.  In designing this observing strategy it is important to
forecast which fraction of all GRBs originate from different redshifts. For
example, it would be impractical to search for those very high redshifts
which amount to a fraction smaller than $10^{-3}$ of all GRBs, because
barely a single one of them would be found by {\it Swift} over several
years of operation.

In this paper, we use existing observational and theoretical work on the
cosmic star formation history to predict the fraction of all GRBs that are
expected to originate at different redshifts. In order to keep our results
general, we make predictions about {\it all} GRBs without reference to the
detection threshold or redshift horizon of any particular instrument.  To
ascertain, however, what the BATSE and {\it Swift} instruments are expected
to detect, we in addition estimate the redshift distributions for these
flux-limited surveys.

In \S 2, we calculate the collapsed fraction of baryons as a function of
redshift based on the Press-Schechter formalism, and infer the
corresponding redshift distribution of GRBs.  In \S 3 we use the inferred
redshift distribution of GRBs to convert the observed distribution of GRB
durations into the corresponding intrinsic distribution, under the simple
assumption that its normalized form is redshift independent. Finally, we
discuss the implications of our results in \S 4.

\section{Structure Formation Model}
\subsection{Star Formation History}

We adopt the popular view that the formation of cosmic structure has
progressed hierarchically from small to large scales, according to a
variant of the cold dark matter (CDM) model. Specifically, we assume a
$\Lambda$CDM model with density parameters in matter
$\Omega_{m}=1-\Omega_{\Lambda}=0.3$ and in baryons $\Omega_{\rm B}=0.045$;
a Hubble constant of $h=H_{0}/(100 {\rm km\,} {\rm s}^{-1} {\rm
Mpc}^{-1})=0.7$, and a scale-invariant power spectrum of density
fluctuations with an amplitude $\sigma_{8}=0.9$ on a scale of $8 h^{-1}{\rm
Mpc}$.

\begin{center} % fig.1
\epsfig{file=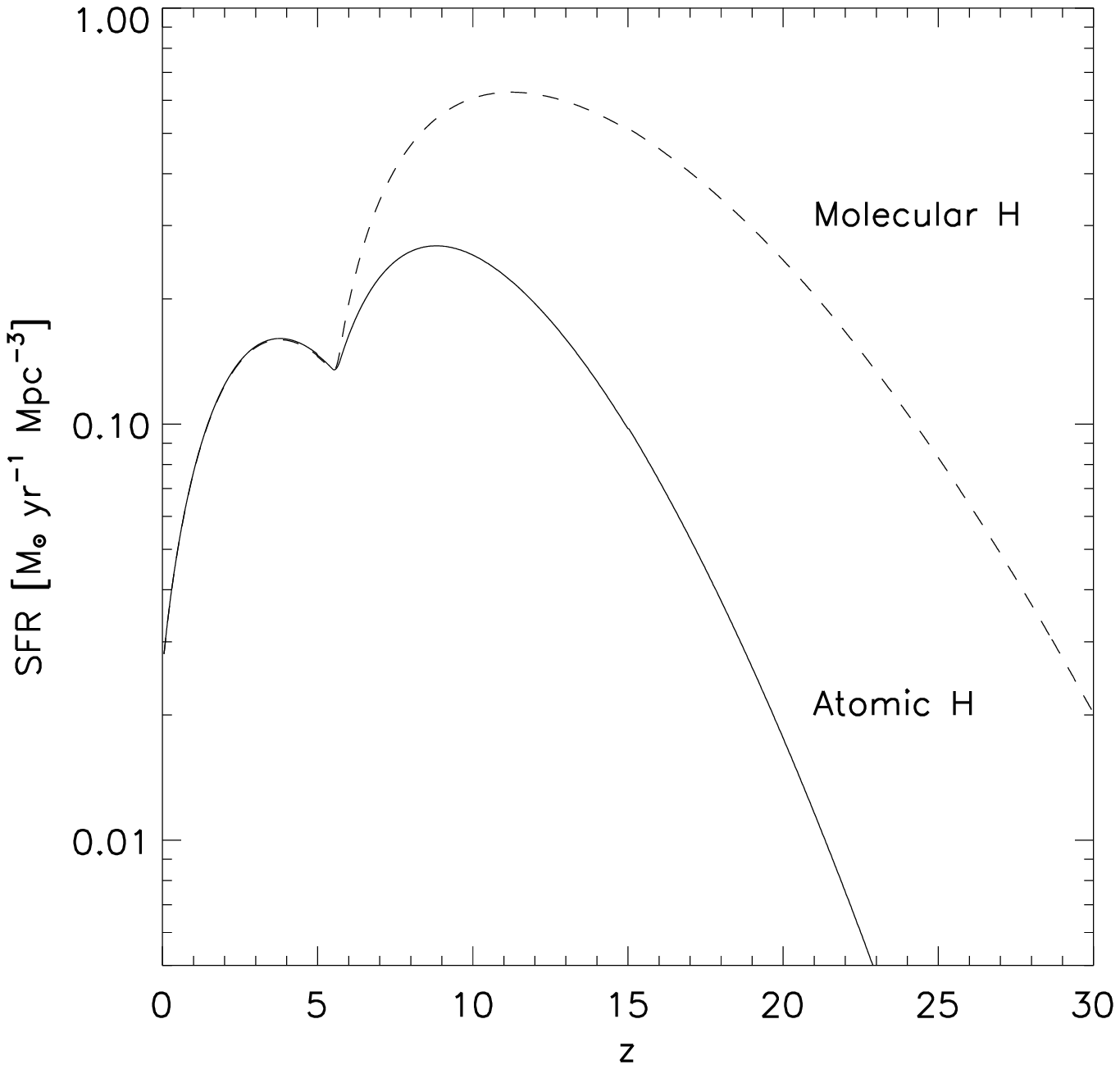,width=8.4cm,height=7.56cm}
\figcaption{History of cosmic comoving star formation rate (SFR) 
in units of $M_{\odot}
\mbox{yr}^{-1} \mbox{Mpc}^{-1}$ as a function of redshift.  {\it Solid
line:} cooling due to atomic hydrogen only; {\it Dashed line:} added
cooling via molecular hydrogen.  The star formation efficiency is assumed
to be $\eta_\ast=10\%$.
\label{fig1}}
\end{center} % fig.1

Our star formation model closely follows that of Barkana \& Loeb (2000),
and here we only briefly describe the key assumptions (see also Santos,
Bromm, \& Kamionkowski 2002). The abundance and merger history of the CDM
halos is described by the extended Press-Schechter formalism (Lacey \& Cole
1993). We assume that the IGM has a two-phase structure, consisting of a
neutral and an ionized phase. The fraction of the cosmic volume filled with
H~II regions is given by
\begin{equation}
f_{\rm ion} = \left\{
\begin{array}{ll}
\exp[-0.495(z-5.6)] & \mbox{for $z > 5.6$}\\
1 & \mbox{otherwise}\\
\end{array}
\right. \mbox{\ .}
\end{equation}
This assumed ionization history fits the semi-analytical calculation of
Barkana \& Loeb (2001) and is consistent with numerical simulations of
reionization (Gnedin 2000, 2001; Razoumov et al. 2001) and the latest data
on quasars in the redshift interval $5\la z\la 6.3$ (Becker et al. 2001;
Djorgovski et al. 2001a; Fan et al. 2001).  At high redshifts, $z\ga 20$,
the universe is predominantly neutral. Once the first luminous objects
form, an increasing fraction of the IGM becomes ionized. At $z_{\rm
reion}\approx 7$, the ionized phase in our model comprises a volume
fraction of $\sim 50$\%, and reionization of the IGM is complete by
$z\approx 5.6$.

Within each phase of the IGM, stars are able to form in two different ways.
The first mechanism pertains to primordial, metal-free, gas. Such gas
undergoes star formation provided that it falls into a sufficiently deep
CDM potential well, or equivalently, into a CDM halo more massive than a
critical value. For the neutral medium, this minimum mass is set by the
requirement for the gas to cool.  Radiative cooling by molecular hydrogen
(H$_2$) allows star formation in halos with a virial temperature $T_{\rm
vir}\ga 300$ K, while atomic cooling dominates for halos with $T_{\rm
vir}\ga 10^{3.9}$ K. The corresponding minimum circular velocities are
$v_{c}\sim 2.5$ km s$^{-1}$ and $\sim 12$ km s$^{-1}$, respectively.  Since
H$_2$ can be easily photo-dissociated by photons below the Lyman-limit, its
significance in the cosmic star formation history is unclear (e.g. Haiman,
Abel, \& Rees 2000; Ricotti, Gnedin, \& Shull 2001, and references
therein), and so we show results with and without H$_2$ cooling. These two
theoretical models are likely to provide conservative bounds for the true
star formation history at $z\ga 5$. The construction of more tightly
constrained models has to await further advances in our understanding, both
observational and theoretical, of star formation at the highest redshifts.

For the ionized medium, on the other hand, the minimum threshold mass is
given by the Jeans mass, since the infall of gas and the subsequent
formation of stars requires that the gravitational force of the collapsing
CDM halo be greater than the opposing pressure force on the gas. After
reionization, the IGM is photo-heated to temperatures $\ga 10^{4}$K,
leading to a dramatic increase in the Jeans mass. We model the suppression
of gas infall according to results from spherically-symmetric collapse
simulations (Thoul \& Weinberg 1996).  Expressing the Jeans mass as the
equivalent halo circular velocity, we assume complete suppression for halos
with $v_{c}\la 35$ km s$^{-1}$, no suppression for $v_{c}\ga 93$ km
s$^{-1}$, and a linear interpolation in between so that $\sim$ 50\%
suppression occurs at $v_{c}\sim 55$ km s$^{-1}$.

Within our model, the second mechanism to form stars occurs in gas that has
experienced a previous burst of star formation, and is therefore already
somewhat enriched with heavy elements.  Such gas, residing in a halo of
mass $M_{1}$, can undergo induced star formation triggered by a merger with
a sufficiently massive companion halo of mass $M_{2}>0.5 M_{1}$.  We
finally assume that stars form with an efficiency of $\eta_\ast\sim 10\%$,
independent of redshift and regardless of whether the gas is primordial or
pre-enriched. This efficiency yields roughly the correct fraction of
$\Omega_{\rm B}$ found in stars in the present-day universe.

Figure 1 shows the resulting star formation histories. Our theoretical
models agree well with observational estimates of the cosmic star formation
rate (SFR) at $z\la 2$ (e.g., Blain et al. 1999).  It is evident that there
are two distinct epochs of cosmic star formation, one at $z\sim 3$, and a
second one at $z\sim 8$ (or at even higher redshifts if H$_{2}$ cooling is
effective). Again, we emphasize that the true history of the cosmic SFR is
likely to lie between the two curves in Figure 1, depending on how complete
the destruction of H$_{2}$ as a function of redshift is.

In deriving the redshift distribution of GRBs in the next section, we do
not make any assumptions on the possible variation of the IMF for stars
forming at different redshifts. Instead, we only assume that baryons are
incorporated into stars, regardless of their specific properties, with the
overall rate calculated in this section. Let us, however, briefly discuss
the possible differences in star formation at high and low redshifts, based
on recent theoretical work implying that star formation at high redshifts
might have proceeded very differently from the present-day case, leading to
stars with typical masses of $M_{\ast}\ga 100 M_{\odot}$ (Bromm et
al. 1999, 2002; Abel et al. 2000, 2002; Nakamura \& Umemura 2001).  After
the first stars have formed, the subsequent generation of stars forms out
of gas that was already enriched with heavy elements. This enriched gas
could have cooled more efficiently, and was able to reach lower
temperatures. Star formation, then, is expected to result in a less
top-heavy IMF. As shown by Bromm et al. (2001), the transition from a
top-heavy to the more standard (Salpeter) IMF occurs when the mass fraction
in metals exceeds a critical value of $\sim 10^{-3}Z_{\odot}$. Gas with a
metal abundance below this threshold is therefore still expected to form
very massive stars. An IGM metal abundance of $\sim 10^{-3}Z_{\odot}$
approximately corresponds to the production of enough ionizing stellar
photons to reionize the universe. Star formation at $z\ga 7$ might
consequently have been dominated by very massive stars, whereas at lower
redshifts, stars form with an IMF close to the Salpeter form.

\subsection{Redshift Distribution of GRBs}

Under the assumption that the formation of GRBs follows closely the cosmic
star formation history with no cosmologically--significant time delay, we
write for the number of GRB events per comoving volume per time: $\psi_{\rm
GRB}(z)=\eta_{\rm GRB}\times \psi_{\ast}(z)$, where $\psi_{\ast}(z)$ is the
stellar mass produced on average per comoving volume per time, as
calculated in \S 2.1. The efficiency factor, $\eta_{\rm GRB}$, links the
formation of stars to that of GRBs, and is in principle a function of
redshift as well as the properties of the underlying stellar population.
%Regarding the latter, by coincidence both a top-heavy IMF in
%the case of Population~III stars and a Salpeter IMF in the case of
%Population~I stars lead to the production of roughly one supernova event
%per $100 M_{\odot}$ in stars. For Population~III stars, this is simply due
%to their large mass so that all of them explode as supernova. The two
%stellar populations differ fundamentally in the way they undergo a
%supernova explosion. Normal, massive Population~I stars end their lives as
%a core collapse supernova, whereas massive Population~III stars give rise
%to pair-instability supernovae, or collapse completely into a massive BH
%(Fryer, Woosley, \& Heger 2001; see also Schneider et al. 2001).  
Massive stars of Population~I differ fundamentally from those of
Population~III; moreover, it is at present not well understood how a
massive Population~III star may give rise to a GRB (see Fryer, Woosley, \&
Heger 2001; Schneider et al. 2001).  Given the current state of uncertainty
with regard to the central engine of GRBs (e.g., Piran 2000), we make the
simplifying assumption that both populations of massive stars are connected
to GRBs in a similar way, and take $\eta_{\rm GRB}$ to be independent of
redshift.  While this simplifying assumption follows from the lack of
better information, our analysis provides a starting point for future
improvements as soon as better observational constraints on high redshift
GRB and star formation will become available.

If we now consider a time interval $\Delta t_{\rm obs}$ in the observer
frame, the total number of GRBs, regardless of whether they are
actually observed or missed, can be written as
\begin{equation}
N(>z)=\int_{z}^{\infty}\psi_{\rm GRB}(z')\frac{\Delta t_{\rm obs}}{(1+z')}
\frac{{\rm d}V}{{\rm d}z'} {\rm d}z' \mbox{\ \ ,}
\end{equation}
where ${\rm d}V/{\rm d}z$ is the comoving volume element per unit redshift,
given by
\begin{equation}
\frac{{\rm d}V}{{\rm d}z}=\frac{4\pi c d_{L}^{2}}{1+z}
\left|\frac{{\rm d}t}{{\rm d}z}\right| \mbox{\ \ .}
\end{equation}
The luminosity distance, $d_{L}$, to a source at redshift $z$ is
\begin{equation}
d_{L}=c(1+z)\int_{0}^{z}(1+z^{\prime})
\left|\frac{{\rm d}t}{{\rm d}z^{\prime}}\right| {\rm d}z^{\prime} \mbox{\ \ ,}
\end{equation}
with
\begin{equation}
\left|\frac{{\rm d}t}{{\rm d}z}\right|^{-1}= (1+z) H_{0}
\sqrt{\Omega_{m}(1+z)^{3}+\Omega_{\Lambda}} 
\end{equation}
in a flat universe.  The fraction of bursts that originate from a redshift
of $z$ or higher, $f(>z)=N(>z)/N(>0)$, is independent of constant
parameters such as $\Delta t_{\rm obs}$, $\eta_{\rm GRB}$ or the beaming
factor of the GRB emission.  The integrand in equation (2) contains the
differential comoving volume element, ${\rm d}V/{\rm d}z$, as is
appropriate for the calculation of an {\it event} rate. Since GRB events
are communicated via photons, we integrate over the redshift-dependent
comoving volume element along our past light cone. We observe these events
over a fixed time window, $\Delta t_{\rm obs}$, which corresponds to
$\Delta t_{\rm obs}/(1+z)$ in the source frame.  If, on the other hand, we
were interested in determining the amount of stellar fossils that have
accumulated over cosmic time in a local comoving volume (see below), we
would have to simply integrate over cosmic time along our past worldline.
\begin{center} % fig.2
\epsfig{file=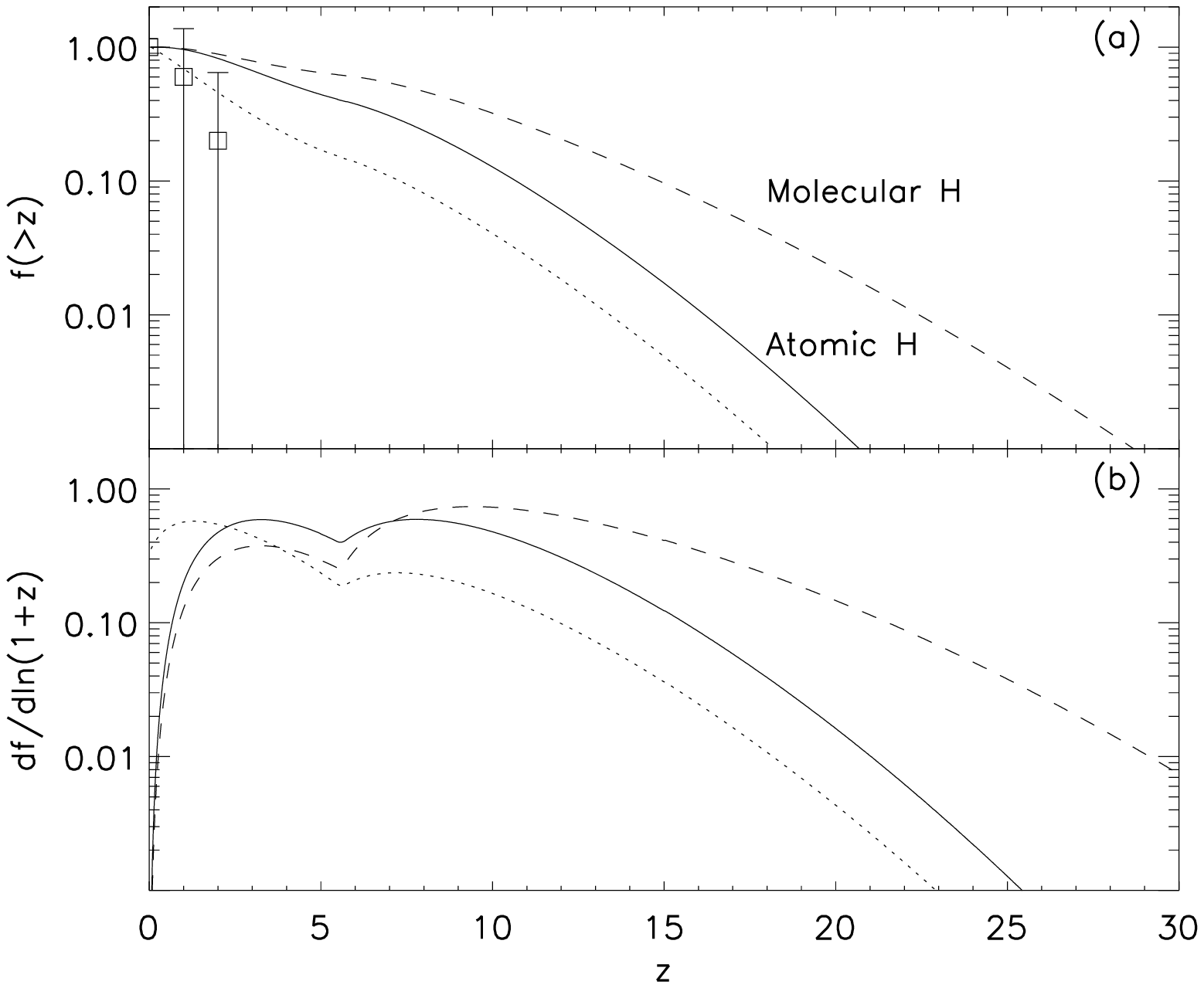,width=8.4cm,height=7.56cm}
\figcaption{Redshift
distribution of GRBs.  ({\it a}) Fraction of bursts that originate at a
redshift higher than $z$ vs. $z$. The curves correspond to the two star
formation histories in Figure 1. The data points reflect $\sim 20$ observed
redshifts (from Ghisellini 2001).  ({\it b}) Fraction of bursts per
logaritmic interval of $(1+z)$ vs. $z$.  The curves have the same meaning
as in panel ({\it a}).  
{\it Dotted lines}: Redshift distribution of the baryonic mass fraction in
stars, calculated for the case of atomic hydrogen cooling.
\label{fig2}}
\end{center} % fig.2

\begin{center} % fig.3
\epsfig{file=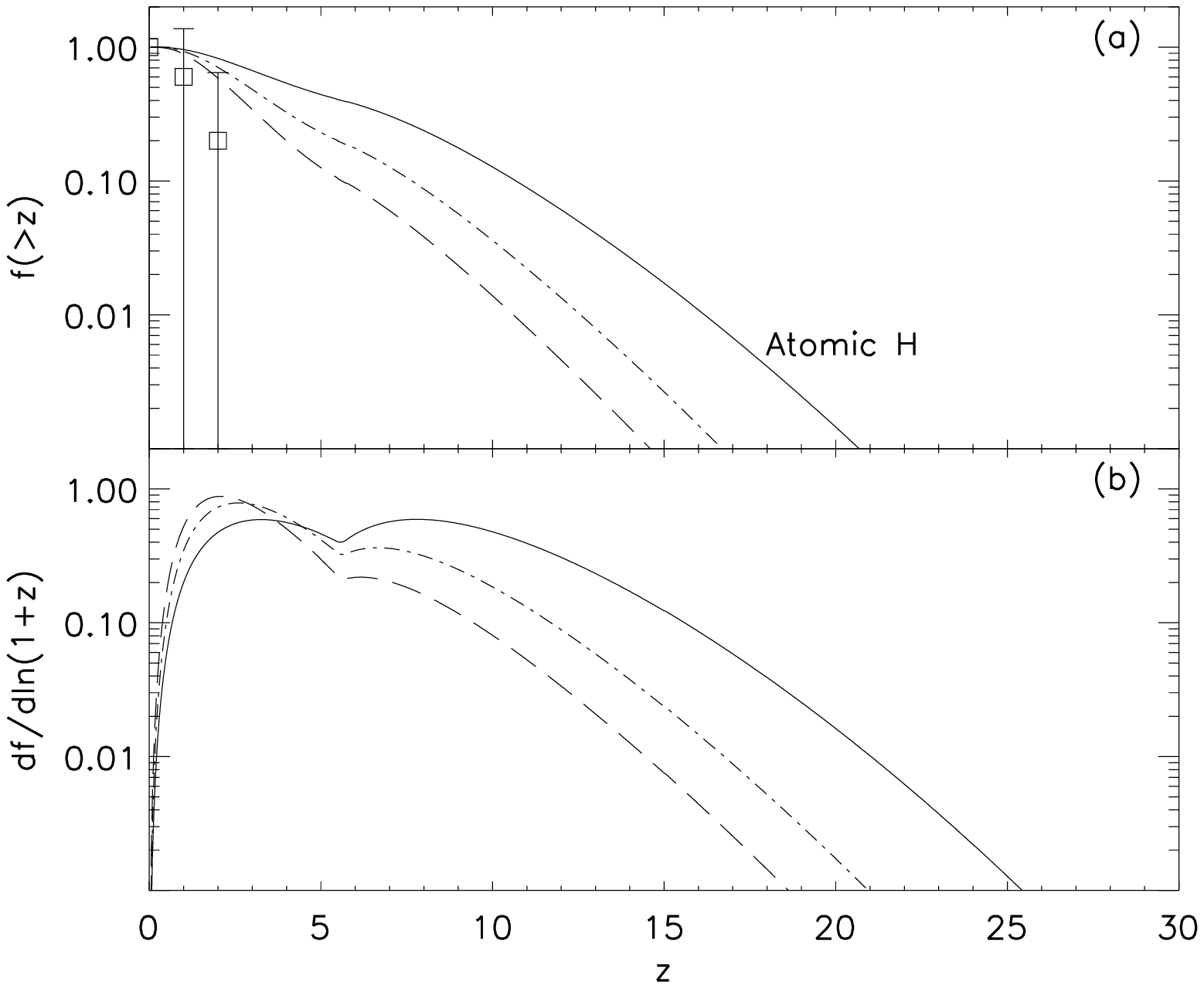,width=8.4cm,height=7.56cm} \figcaption{Redshift
distribution of all GRBs as compared to that measured by flux-limited
surveys.  The convention for the panels and the meaning of the symbols is
the same as in Fig. 2.  {\it Solid lines}: All GRBs for star formation
through atomic line cooling.  {\it Dot-dashed lines}: Expected distribution
for {\it Swift}.  {\it Long-dashed lines}: Expected distribution for BATSE.
Note that the curves for the two flux-limited surveys are very uncertain
due to the poorly-determined GRB LF.
\label{fig3}}
\end{center} % fig.3

In Figure 2, we show $f(>z)$ together with the differential distribution
$({\rm d}f/{\rm d} z)$ for the two star formation histories of Figure 1.
It is evident that a significant fraction of all bursts is predicted to
occur at high redshifts, namely $f(z\ge 5)\ga 50\%$, and that a few percent
of all bursts occur at redshifts as high as $z\sim 20$.  Evaluating the
mean redshift for GRBs using the distributions of Figure 2, we find
$\bar{z}\sim 5$ in the case of atomic cooling and $\bar{z}\sim 8$ for
molecular cooling.  Currently, only $\sim 20$ bursts have known redshifts
(Kulkarni et al. 2000; Djorgovski et al. 2001c; Ghisellini 2001), and we
include this small observed sample in Figure 2. The fact that the data
points lie below our theoretical prediction could be due to small-number
statistics as well as due to a redshift--dependent incompleteness bias.

We stress that equation (2) gives the fraction of transient events observed
on the sky, and not the fraction of all baryons that have been incorporated
into stars by a redshift $z$ relative to the same fraction today. This
latter quantity is given by 
\begin{equation}
f_{\ast}(>z) =\frac{\int_{z}^{\infty}
\psi_{\ast}(z')\left| {\rm d}t/{\rm d}z'\right|{\rm
d}z'}{\int_{0}^{\infty} \psi_{\ast}(z')\left| {\rm d}t/{\rm
d}z'\right|{\rm d}z'} \mbox{\ ,}
\end{equation}
and is shown in Figure 2 for the case of atomic cooling. As can be seen,
the fraction of all stars that are formed at $z\ga 5$ is $\sim 15$\%. {\it
Collecting photons from our past light cone is therefore a more efficient
way of probing the first stars than sorting through the fossil stellar
record in the present-day universe.}

It is important to emphasize that the analysis presented here pertains to
{\it all} bursts, regardless of whether existing or previous instruments
have actually been able to detect them. If the horizon of previous
instruments was limited to $z\ll 5$, then our predictions provide important
motivation for the construction of more-sensitive instruments that would
trigger on GRBs out to the highest redshifts. The fraction of all bursts
that were detected by any given instrument depends on the
instrument-specific flux sensitivity threshold and on the poorly-determined
luminosity function (LF) of GRBs (see e.g., Schaefer, Deng, \& Band 2001;
Schmidt 2001; Norris 2002).

It is nevertheless instructive to ascertain what existing or planned
instruments like BATSE and {\it Swift} are expected to find. To this
extent, we modify the GRB event rate to have
\begin{equation}
\psi_{\rm GRB}(z)=\eta_{\rm GRB} \psi_{\ast}(z)
\int_{L_{\rm lim}(z)}^{\infty}p(L) {\rm d}L \mbox{\ \ .}
\end{equation}
Here, $p(L)$ is the GRB LF with $L$ being the intrinsic photon luminosity
(in units of photons s$^{-1}$). If $f_{\rm lim}$ denotes the sensitivity
threshold of a given instrument (in photons s$^{-1}$ cm$^{-2}$), then the
minimum luminosity is
\begin{equation}
L_{\rm lim}(z)= 4 \pi d_{L}^{2} f_{\rm lim} \mbox{\ \ .}
\end{equation}
This expression is derived with a spectral index of $\alpha=2$ for $L\propto
\nu^{-\alpha}$ (Band et al. 1993). For definiteness, we assume
a log-normal distribution function (e.g., Woods \& Loeb 1995)
\begin{equation}
p(L)= \frac{{\rm e}^{-\sigma^{2}/2}}{\sqrt{2\pi \sigma^{2}}}
\exp [-(\ln(L/L_{0}))^{2}/(2\sigma^{2})]\frac{1}{L_{0}} \mbox{\ \ ,}
\end{equation}
where $\sigma$ and $L_{0}$ are the width and the average luminosity,
respectively. Recently, Sethi \& Bhargavi (2001) have shown that both the
observed number count--flux relation as well as the existing afterglow
redshift data are consistent with a log-normal LF for best-fit parameters
(taking into account the effect of beaming): $\sigma=2$ and $L_{0}=2\times
10^{56}$s$^{-1}$, and we adopt these values in the following analysis. To
determine the expected redshift distribution as observed by BATSE and {\it
Swift}, we use equation (2) together with the GRB rate given in equation
(7). The flux thresholds are $f_{\rm lim}=0.2$ and 0.04 photons s$^{-1}$
cm$^{-2}$ for BATSE and {\it Swift}, respectively (Lamb \& Reichart 2000,
and references therein). In Figure 3, we show the same quantities as in
Figure 2, but now comparing the distributions for BATSE and {\it Swift}
with our theoretical prediction for atomic line cooling. It can be seen
that in the case of BATSE a fraction of $f(z\ge 5)\ga 10$\% of all bursts
originates from high redshifts, whereas the corresponding fraction for {\it
Swift} is $f(z\ge 5)\ga 25$\%. We emphasize again that these numbers are
uncertain due to the poorly-known GRB LF. Figure 3 nicely demonstrates the
asymptotic character of our theoretical prediction, pertaining to a future
`ultimately-sensitive' instrument. Indeed, using the LF above, we estimate
that an instrument with a sensitivity of $\sim 50$ times better than {\it
Swift} would be able to detect the full theoretically-possible sample of
bursts from $z\ga 5$.

The detectability of $z\ga 5$ GRBs is also a crucial ingredient in
estimating the fraction of all well-localized bursts that have no
detectable optical afterglow, the so-called ``dark GRBs''. Various authors
have used the fraction of dark bursts in the currently observed sample of
GRBs to constrain the amount of dust obscured star formation (e.g.,
Djorgovski et al. 2001b). The resulting fraction of dark GRBs estimated for
different redshifts depends on the, presently unknown, level of
incompleteness in the observed sample.  In the context of our model, we
predict that {\it all} GRB afterglows originating at $z\ga 6$ are optically
dark.  The intervening, partially neutral IGM would efficiently absorb the
rest-frame UV afterglow that would otherwise have been redshifted into the
optical band (see also Fruchter 1999; Piro et al. 2002).  These bursts
might give rise to the recently discovered class of X-ray rich GRBs (e.g.,
Piro et al. 2002; see also Schneider et al. 2001) due to the redshifting of
the source-frame $\gamma$-rays into the X-ray band.

\section{Burst Durations} 

The duration of a GRB reflects the characteristic timescale over which the
central engine is active and is therefore a diagnostic of the GRB
progenitor. The distribution of GRB durations has been determined by the
BATSE instrument on board the {\it Compton Gamma Ray Observatory} (as
summarized in Paciesas et al. 1999), and is observed to be bimodal with a
population of short bursts centered on $T_{\rm obs}\sim 0.3 {\rm \,s}$, and
long bursts around $T_{\rm obs}\sim 30 {\rm \,s}$ (Kouveliotou et
al. 1993). For the definition of the burst duration, $T_{\rm obs}$, we use
the interval of time over which a GRB contains from 5 to 95\% of its total
observed $\gamma$-ray counts, also denoted as $T_{90}$ in the
literature. Since bursts originate over a broad range of redshifts, the
question arises as to what the intrinsic distribution of durations is
like. For simplicity, we assume in this section that BATSE was capable of
sampling the full redshift distribution of GRBs shown in Figure 2. This
provides us with the maximum level of distortion that cosmological time
dilation could have had on the observed distribution of burst durations.
The true intrinsic distribution of durations for the BATSE-triggered bursts
should lie in between the observed distribution and the intrinsic one
calculated in this section.

The number of observed bursts in a given bin $i$, with an observed duration
$T_{{\rm obs,}i}$ and a width $\Delta T_{{\rm obs,}i}$, can be written as
\begin{equation}
\Delta N_{{\rm obs,}i}=N_{\rm tot} \Delta T_{{\rm obs,}i}
\int_{0}^{\infty}\frac{{\rm d}P}{{\rm d}T}(T)\frac{1}{(1+z)}
\frac{{\rm d}f}{{\rm d}z}(z) {\rm d}z \mbox{\ ,}
\end{equation}
where $N_{\rm tot}=\sum_{i} \Delta N_{{\rm obs,}i}$ is the total number of
bursts in the sample.  We assume that the intrinsic distribution, ${\rm
d}P/{\rm d}T$, is independent of redshift and satisfies
$\int_{0}^{\infty}({{\rm d}P}/{{\rm d}T}){\rm d}T=1$ and $({{\rm d}P}/{{\rm
d}T}) \ge 0$.  The intrinsic burst duration, $T$, is related to the
observed one by the cosmological time dilation, $T=T_{{\rm obs,}i}/(1+z)$,
and $({\rm d}f/{\rm d}z)$ is the GRB redshift distribution, as calculated
in \S 2.  We replace the integration in equation (10) by a summation,
covering the range of intrinsic durations, $T$, with the same number of
bins, $N_{\rm bin}=23$, as the observed histogram.  The inversion problem
is then uniquely defined. We carry out the deconvolution with the standard
iterative Lucy method. This is a reliable technique, derived from Bayes'
theorem, to solve a set of linear equations with additional constraints on
the unknowns (Lucy 1974). The stability of the algorithm is improved by
limiting the change in the unknowns in each iteration, and by smoothing
over adjacent bins. To this extent, we use equation (11b) in Baugh \&
Efstathiou (1993) with parameter values of $\beta=0.8$ and $\epsilon=0.9$.
We have verified that the solutions are not very sensitive to the choice of
these parameters.

The result of this inversion is shown in Figure 4, where we compare the
derived intrinsic distribution to the observed one, ${\rm d}P/{\rm d}T_{\rm
obs}(T_{{\rm obs,}i})=\Delta N_{{\rm obs,}i}/ (N_{\rm tot}\Delta T_{{\rm
obs,}i})$. It is evident that the intrinsic durations are systematically
shifted to shorter values due to the cosmological time dilation. The
bimodality is preserved, with peaks that are narrower than the observed
ones (note that the horizontal scale is logarithmic). The two star
formation histories discussed in \S 2 lead to similar intrinsic
distributions.  The shift to shorter durations, however, is more pronounced
in the case of star formation via H$_{2}$ cooling. The mean intrinsic
durations characterising the first, short-duration, peak are $\sim 0.05{\rm
\,s}$ for cooling due to atomic hydrogen, and $\sim 0.03{\rm \,s}$ for
H$_{2}$ cooling. The corresponding numbers for the long-duration peak are
$\sim 7{\rm \,s}$ and $\sim 5{\rm \,s}$, respectively. These differences in
the mean durations are a direct consequence of the fact that GRBs originate
on average at somewhat higher redshift if H$_{2}$ cooling is effective.
Note that, statistically, the longest duration bursts, with $T_{\rm obs}\ga
1000{\rm \, s}$, are expected to originate at high $z$, and this could be a
successful selection strategy for observations targeting high--redshift
GRBs.  The shift to longer durations due to the cosmological time dilation
could in part be compensated by the following subtle selection effect which
we ignore in this paper. Sources at high $z$ will on average have lower
fluxes, and observations with a given sensitivity threshold will therefore
only detect the brightest portion of the total emission, thus
systematically underestimating the true duration of the burst.

\begin{center} % fig.3
\epsfig{file=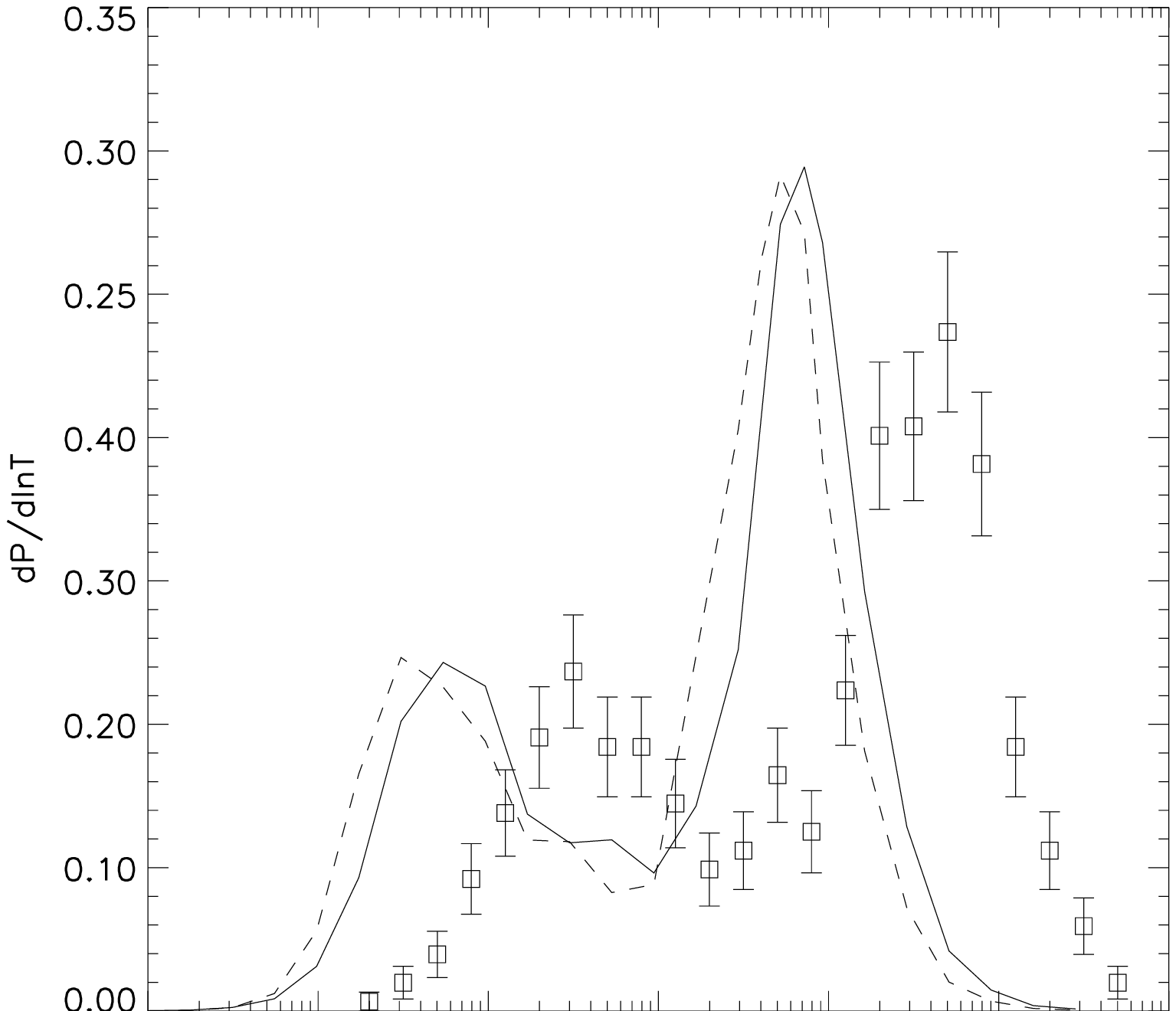,width=8.4cm,height=7.56cm}
\vspace{0.8cm} \figcaption{The maximal effect that cosmological
time-dilation may have on the observed distribution of GRB
durations. Distributions are shown as probability per logarithmic duration
interval vs.  duration (in s).  {\it Solid line:} Case of atomic hydrogen
cooling.  {\it Dashed line:} Case with molecular cooling included as well.
The data points correspond to the observed distribution by BATSE (Paciesas
et al. 1999).  The intrinsic durations are systematically shifted to
smaller values due to cosmological time dilation.
\label{fig4}}
\end{center} % fig.4

Based on observations of GRB afterglows with known redshifts, Frail et
al. (2001) have recently presented evidence for a standard amount of energy
release in GRBs, $E\sim 10^{51}{\rm \,erg}$. Making the simplest assumption
of a constant energy for all long-duration bursts (which are the only ones
with measured redshifts so far), one can easily derive the luminosity
function from the intrinsic distribution of burst durations. The luminosity
of a burst is then simply $\approx E/T$, and the resulting luminosity
function is obtained by inverting the horizontal axis in Figure 4 and
changing $T$ to $E/T$. The long-duration bursts would then narrowly cluster
around a luminosity of $\sim 10^{50}$erg s$^{-1}$.

\section{Discussion}

We have derived the redshift distribution of GRBs out to $z\ga 20$ under
the assumption that the GRB event rate traces the cosmic star formation
rate. We find that $\ga 50$\% of all GRBs on the sky originate from a
redshift of 5 or higher. On the other hand, the fraction of baryons that
have been incorporated into stars by $z\sim 5$ is much smaller, comprising
only $\sim 15$\% of the stellar mass formed by today.  The difference
between the two fractions follows from the different cosmological factors
in the redshift integrations for the statistics of transient events on the
sky as compared to the census of fossil objects in the local universe.  The
favorable statistical bias towards high-redshift events on the sky is
expected to apply also to Type II supernova explosions which are related to
the formation of massive stars in a similar way as GRBs. Despite their
dimming with increasing redshift, high--redshift supernovae will be
detectable with sufficiently sensitive telescopes such as the {\it Next
Generation Space Telescope}\footnote{http://ngst.gsfc.nasa.gov/} (NGST;
Miralda-Escud\'e \& Rees 1997; Woods \& Loeb 1998).  In fact, our
calculation implies that without any additional bias (such as
redshift-dependent dust extinction) approximately half of all Type II
supernovae detected by NGST will originate at $z\ga 5$.  Deep observations
of high--redshift GRBs and supernovae offer an ideal window into the
earliest epoch of cosmic structure formation. The lengthening of the
duration of these transients by a factor $(1+z)$ makes it easier for
observers to monitor their lightcurves.

Different instruments may find GRBs up to different redshifts, depending on
their detection sensitivity and the highly uncertain GRB luminosity
function (Schaefer et al. 2001; Schmidt 2001; Norris 2002). A
trigger-unbiased way to infer the redshift evolution of the GRB event rate
is to compare the number counts of GRBs with the same absolute (intrinsic)
luminosity in different redshift bins.  If future observations of this type
were to determine a mean redshift for the GRB distribution significantly
lower than the one predicted in this paper, then this would indicate either
that GRB formation at high $z$ is substantially suppressed, or that GRBs
originate from the coalescence of binaries with a time delay of a few Gyr
between the formation of a massive star and the GRB event.

Recent observations indicate that a large fraction, $\sim 50$\%, of all
well-localized GRBs have no associated optical afterglow, and are
classified as ``(optically) dark GRBs'' (e.g., Piro et al. 2002).
According to our model, a substantial fraction of these dark bursts could
originate from $z\ga 6$.  The intervening, partially neutral IGM would
efficiently absorb the rest-frame UV afterglow that would otherwise have
been redshifted into the optical band.

\acknowledgements

We thank Rennan Barkana, Jeremy Heyl, Jonathan Mackey, Bohdan Paczy\'{n}ski
and Martin Rees for helpful discussions.  This work was supported in part
by NASA grants NAG 5-7039, 5-7768, and by NSF grants AST-9900877,
AST-0071019.

%%\clearpage
%%\setcounter{figure}{0}
%

\end{document}